\documentclass[aps,prl,amsmath,floatfix,twocolumn,groupaddress]{revtex4}

\usepackage{epsfig}
\usepackage{epsf}
\usepackage{array}
\begin{document}

\bibliographystyle{apsrev}
\preprint{}

\title{Differential Charge Sensing and Charge Delocalization \\
in a Tunable Double Quantum Dot}
\author{L.~DiCarlo, H.~J.~Lynch, A.~C.~Johnson, L.~I.~Childress, K.~Crockett, C.~M.~Marcus}
\affiliation{Department of Physics, Harvard University, Cambridge, Massachusetts 02138}
\author{M.~P.~Hanson, A.~C.~Gossard}
\affiliation{Materials Department, University of California, Santa Barbara, California
93106}

\begin{abstract} We report measurements of a tunable double quantum dot, operating in the quantum regime, with
integrated local charge sensors.  The spatial resolution of the sensors  allows the charge distribution
within the double dot system to be resolved at fixed total charge.  We use this readout scheme to investigate charge
delocalization as a function of temperature and strength of tunnel coupling, demonstrating that local charge sensing can be used to
accurately determine the interdot coupling in the absence of transport.
\end{abstract}

\maketitle

Coupled semiconductor quantum dots have proved a fertile ground
for exploring quantum states of electron charge and spin. These
``artifical molecules" are a scalable technology with possible
applications in information processing, both as classical switching elements \cite{Chan02, Gardelis03} or as charge or spin qubits \cite{Engel03}.
Charge-state superpositions may be probed using tunnel-coupled
quantum dots, which provide a tunable two-level system whose two
key parameters, the bare detuning $\epsilon$ and tunnel coupling
$t$ between two electronic charge states \cite{vanderWiel03}, can
be controlled electrically.

In this Letter, we investigate experimentally a quantum two-level
system, realized as left/right charge states in a gate-defined
GaAs double quantum dot, using local electrostatic sensing (see
Fig.~1). In the absence of tunneling, the states of the two-level
system are denoted $(M+1,N)$ and $(M,N+1)$, where the pair of
integers refers to the number of electrons on the left and right
dots. For these two states, the total electron number is fixed,
with a single excess charge moving from one dot to the other as a
function of gate voltages.  When the dots are tunnel coupled, the
excess charge becomes delocalized and the right/left states
hybridize into symmetric and antisymmetric states.

Local charge sensing is accomplished using integrated quantum
point contacts (QPC's) positioned at opposite sides of the double
dot. We present a model for charge sensing in a tunnel-coupled
two-level system, and find excellent agreement with experiment.
The model allows the sensing signals to be calibrated using
temperature dependence and measurements of various capacitances.
For significant tunnel coupling, $0.5 k_B T_e\lesssim t < \Delta$
($T_e$ is electron temperature, $\Delta$ is the single-particle
level spacing of the individual dots), the tunnel coupling $t$ can
be extracted quantitatively from the charge sensing signal,
providing an improved method for measuring tunneling in quantum
dot two-level systems compared to transport methods
\cite{vanderWiel03}.

\begin{figure}
      \label{Figure1}
  \includegraphics[width=2.5in]{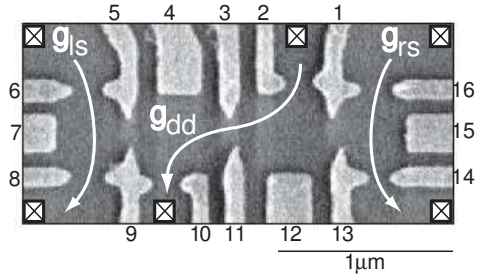}
\caption{ \footnotesize{SEM micrograph of a device similar to the measured device, consisting of a double quantum dot
with quantum point contact charge sensors formed by gates 8/9 (13/14) adjacent to the left (right) dot. Series conductance $g_{\mathrm{dd}}$
through the double dot was measured simultaneously with conductances $g_{\mathrm{ls}}$ and $g_{\mathrm{rs}}$ through the
left and right sensors.}}
\end{figure}

Charge sensing using a QPC was first demonstrated in Ref.~\cite{Field93}, and has been used previously to investigate
charge delocalization in a single dot strongly coupled to a lead in the classical regime \cite{Duncan99}, and as a means
of placing bounds on decoherence in an isolated double quantum dot
\cite{Gardelis03}. The back-action of a QPC sensor, leading to phase decoherence, has been investigated experimentally
\cite{Buks98} and theoretically \cite{backaction}. Charge sensing with sufficient
spatial resolution to detect charge distributions within a double dot has been demonstrated in a metallic system
\cite{Amlani97,Buehler03}. However, in metallic systems the interdot tunnel coupling cannot be tuned, making the
crossover to charge delocalization difficult to investigate. Recently, high-bandwidth charge sensing using a metallic
single-electron transistor \cite{Schoelkopf98}, allowing individual charging events to be counted, has been demonstrated
\cite{Lu03}.
  Recent measurements of gate-defined few-electron GaAs double dots \cite{Elzerman03} have demonstrated dual-QPC charge
sensing down to $N,M = 0,1,2...$, but did not focus on sensing at fixed electron number, or on charge delocalization.
The present experiment uses larger dots, containing $\sim 200$ electrons each (though still with temperature less than
level spacing, see below).

\begin{figure}
      \label{Figure2}
     \includegraphics[width=3.2in]{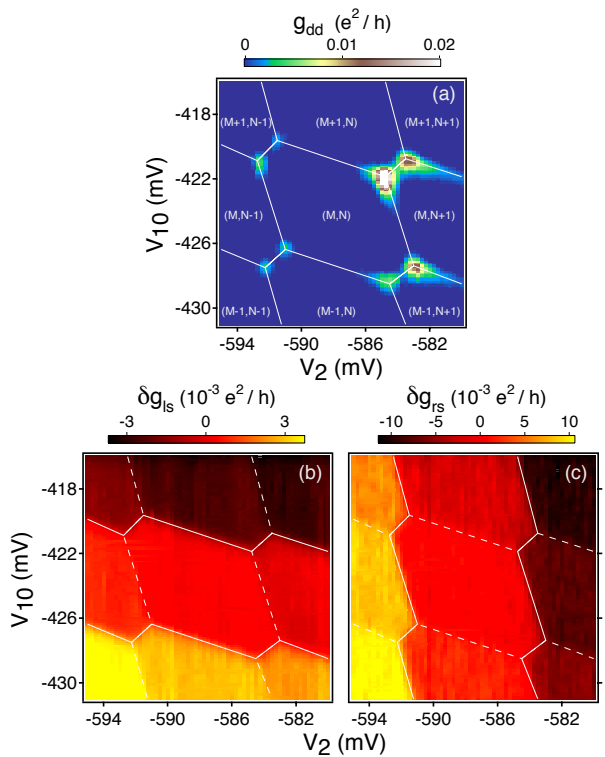}
\caption{ \footnotesize{(a) Double dot conductance,
$g_{\mathrm{dd}}$, as a function of gate voltages $V_{2}$ and
$V_{10}$.  White lines indicate the honeycomb pattern. Within each
honeycomb cell, electron number on each  dot is well defined, with
M(N) referring to electron number in the left (right) dot. (b, c)
Simultaneously measured sensing signals from left (b) and right
(c) QPCs. $\delta g_{\mathrm{ls}}$ ($\delta g_{\mathrm{rs}}$) are
QPC conductances after subtracting a best-fit plane. See text for
details. The horizontal pattern in (b) and vertical pattern in (c)
demonstrate that each sensor is predominantly sensitive to the
charge on the dot it borders.}}
\end{figure}

The device we investigate, a double quantum dot with adjacent
charge sensors, is formed by sixteen electrostatic gates on the
surface of a GaAs/Al$_{0.3}$Ga$_{0.7}$As heterostructure grown by
molecular beam epitaxy (see Fig.~1). The two-dimensional electron
gas layer, 100 nm below the surface, has an electron density of
$2\times10^{11}\, \mathrm{cm}^{-2}$ and mobility $2\times10^{5}\,
\mathrm{cm}^2/\mathrm{Vs}$. Gates 3/11 control the interdot tunnel
coupling while gates 1/2 and 9/10 control coupling to electron
reservoirs. In this measurement, the left and right sensors were
QPCs defined by gates 8/9 and 13/14, respectively; gates 6, 7, 15,
and 16 were not energized.  Gaps between gates 5/9 and 1/13 were
fully depleted, allowing only capacitive coupling between the
double dot and the sensors.

Series conductance, $g_{\mathrm{dd}}$,  through the double dot was measured using standard lock-in techniques with a
voltage bias of $5\, \mu\mathrm{V}$ at 87 $\mathrm{Hz}$. Simultaneously, conductances  through the left and right QPC
sensors, $g_{\mathrm{ls}}$ and $g_{\mathrm{rs}}$, were measured in a current bias configuration using separate lock-in
amplifiers with $0.5\, \mathrm{nA}$ excitation at 137 and 187 $\mathrm{Hz}$. Throughout the experiment, QPC sensor
conductances were set to values in the 0.1 to 0.4 $e^2/h$ range by adjusting the voltage on gates 8 and 14.

The device was cooled in a dilution refrigerator with base
temperature $T \sim30\, \mathrm{mK}$.  Electron temperature $T_e$
at base was $\sim 100\, \mathrm{mK}$, measured using Coulomb
blockade peak widths with a single dot  formed. Single-particle
level spacing $\Delta \sim\, 80\, \mu \mathrm{eV}$ for the
individual dots was also measured in a single-dot configuration
using differential conductance measurements at finite drain-source
bias. Single-dot charging energies, $E_C=e^2/C_o \sim 500\, \mu
\mathrm{eV}$ for both dots (giving dot capacitances  $C_o\sim320
\, \mathrm{aF}$), were extracted from the height in bias of
Coulomb blockade diamonds \cite{LK97}.

 Figure~2(a) shows $g_{\mathrm{dd}}$ as a function of gate voltages $V_{2}$ and $V_{10}$,
exhibiting the familiar `honeycomb' pattern of series conductance
through tunnel-coupled quantum dots \cite{Pothier92,Blick96,
Livermore96}. Conductance peaks at the honeycomb vertices, the
so-called triple points, result from simultaneous alignment of
energy levels in the two dots with the chemical potential of the
leads. Although conductance can be finite along the honeycomb
edges as a result of cotunneling, here it is suppressed by keeping
the dots weakly coupled to the leads. Inside a honeycomb, electron
number in each dot is well defined as a result of Coulomb
blockade. Increasing $V_{10}$ ($V_{2}$) at fixed $V_{2}$
($V_{10}$) raises the electron number in the left (right) dot one
by one.

Figures~2(b) and 2(c) show left and right QPC sensor signals
measured simultaneously with $g_{\mathrm{dd}}$. The sensor data
plotted are $\delta g_{\mathrm{ls(rs)}}$, the left (right) QPC
conductances after subtracting a best-fit plane (fit to the
central hexagon) to remove the background slope due to
cross-coupling of the plunger gates (gates 2 and 10) to the QPCs.
The left sensor shows conductance steps of size $\sim
3\times10^{-3}\, e^2/h$ along the (more horizontal) honeycomb
edges where the electron number on the left dot changes by one
(solid lines in Fig.~2(b)); the right sensor shows conductance
steps of size $\sim1\times10^{-2}\, e^2/h$ along the (more
vertical) honeycomb edges where the electron number of the right
dot changes by one (solid lines in Fig.~2(c)). Both detectors show
a conductance step, one upward and the other downward, along the
$\sim 45$-degree diagonal segments connecting nearest triple
points. It is along this shorter segment that the total electron
number is fixed; crossing the line marks the transition from
$(M+1,N)$ to $(M,N+1)$. Overall, we see that the transfer of one
electron between one dot and the leads is detected principally by
the sensor nearest to that dot, while the transfer of one electron
between the dots is detected by both sensors, as an upward step in
one and a downward step in the other, as expected.

\begin{figure}
      \label{Figure3}
     \includegraphics[width=3.2in]{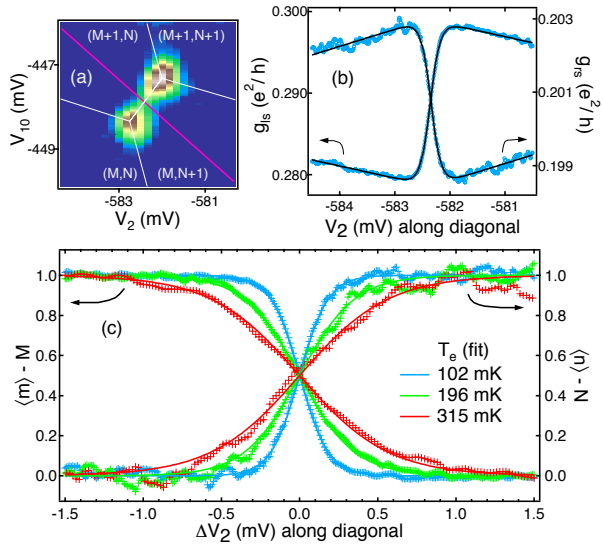}
\caption{ \footnotesize{(a) Double dot conductance
$g_{\mathrm{dd}}$ as a function of gate voltages $V_{2}$ and
$V_{10}$ in the vicinity of a triple point. Same color scale as in
Fig.~1(a). The detuning diagonal (red line) indicates the
fixed-charge transition between $(M+1,N)$ and $(M,N+1)$. (b) Left
and right QPC conductance with no background subtraction (blue
points), along the detuning diagonal, with fits to the two-level
model, Eq.~(2) (black curves). See text for fit details. (c)
Excess charge (in units of $e$) in the left and right dot, at $T =
30\, \mathrm{mK}$ (blue), $200\, \mathrm{mK}$ (green) and $315\,
\mathrm{mK}$ (red). Corresponding values of $T_e$ extracted from
the fits (solid curves) are $102$, $196$ and $315 \,
\mathrm{mK}$.}}
\end{figure}

\begin{figure}
      \label{Figure4}
     \includegraphics[width=3.2in]{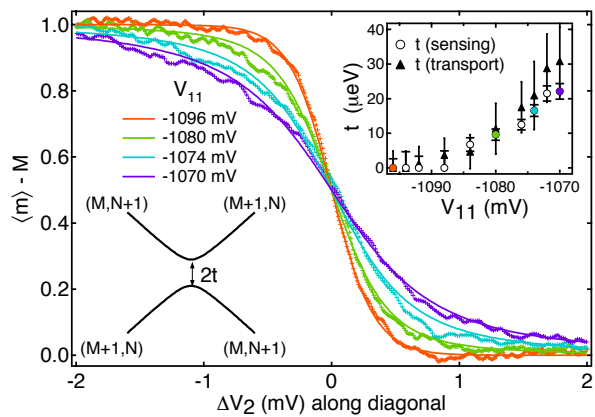}
\caption{\footnotesize{Excess charge on the left dot, extracted
from left QPC conductance data, along a detuning diagonal
(crossing different triple points from those in Fig.~3) at base
temperature and several settings of the coupling gate 11. The
temperature-broadened curve (red) widens as $V_{11}$ is made less
negative, increasing the tunnel coupling, $t$. See text for
details of fits (solid curves). Top right inset: comparison of $t$
values extracted from sensing (circles) and transport (triangles)
measurements, as a function of $V_{11}$. Colored circles
correspond to the transitions shown in the main graph. Lower left
inset: Schematic energy diagram of the two-level system model,
showing ground and excited states as a function of detuning
$\epsilon$, with splitting (anticrossing) of $2t$ at
$\epsilon=0$.}}
\end{figure}

Focusing on interdot transitions at fixed total charge, i.e.,
transitions from $(M+1,N)$ to $(M,N+1)$,  we present
charge-sensing data taken along the ``detuning" diagonal by
controlling gates $V_2$ and $V_{10}$,  shown as a red diagonal
line between the triple points in Fig.~3(a). Raw data (no
background subtracted) for the two sensors are shown in Fig.~3(b).
The transfer of the excess charge from left dot to right dot
causes conductance steps on both QPCs, clearly discernable from
 background slopes caused by coupling of gates 2 and 10 to the
QPCs.

Also shown in Fig.~3(b) are fits to the raw sensor data based on a
model of local sensing of an isolated two-level system in thermal
equilibrium, which we now describe. Varying  $V_2$ and $V_{10}$
along the red diagonal changes the electrostatic energy
difference, or bare detuning $\epsilon$, between $(M+1,N)$ and
$(M,N+1)$ states.  The lever arm relating gate voltage to detuning is set
by the slope of the diagonal cut (see Fig.~3(a)) and various dot
capacitances, and can be calibrated experimentally as described
below. When the tunnel coupling $t$ mixing these two states is
small compared to the single-particle level spacings for the
individual dots, we can consider a two-level system whose ground
and excited states, separated by an energy
$\Omega=\sqrt{\epsilon^2+ 4t^2}$, consist of superpositions of
$(M+1,N)$ and $(M,N+1)$ \cite{Cohen96}. The probability of finding
the excess charge on the left dot while in the ground (excited)
state is $\frac{1}{2}(1\mp\epsilon/\Omega)$. The excited state is
populated at finite temperature, with an average occupation
$1/(1+\exp(-\Omega/k_BT_e))$. The average excess charge (in units
of $e$) on the left and right dots is thus:
\begin{equation}
\left\{
\begin{array}{c}
\langle m \rangle -M \\
\langle n \rangle -N
\end{array}\right\}
= \frac{1}{2}\left(1 \mp \frac{\epsilon}{\Omega}
\mathrm{tanh}\left(\frac{\Omega}{2 k_B T_e}\right)\right).
\end{equation}
 Our model assumes that each sensor responds linearly to the average excess
charge on each dot, but more sensitively to that on the nearest
dot as demonstrated experimentally in Fig.~2. The resulting model
for sensor conductance is:
\begin{equation}
g_{\mathrm{ls(rs)}}=g_{o\mathrm{l}(o\mathrm{r})} \pm \delta
g_{\mathrm{l(r)}} \frac{\epsilon}{\Omega}
\mathrm{tanh}\left(\frac{\Omega}{2 k_B T_e}\right) + \frac{
\partial g_{\mathrm{l(r)}}}{
\partial \epsilon}\epsilon.
\end{equation}
The first term on the right is the background conductance of the
QPC, the second term represents the linear response to average
excess charge, and  the third represents direct coupling of the
swept gates to the QPC. As shown in Fig.~3(b), our model gives
very good fits to the data. For each trace (left and right
sensors), fit parameters are $g_{o\mathrm{l}(o\mathrm{r})}$,
$\delta g_{\mathrm{l(r)}}$, $\frac{\partial
g_{\mathrm{l(r)}}}{\partial \epsilon}$, and $T_e$. In these data,
the tunnel coupling is weak, and we may set $t = 0$.

Figure~3(c) shows the effect of increasing electron temperature on
the transition width. Here, vertical axes show excess charge
extracted from fits to QPC sensor conductance data. Sweeps along
the red diagonal were taken at refrigerator temperatures of $30\,
\mathrm{mK}$ (blue), $200\, \mathrm{mK}$ (green) and $315\,
\mathrm{mK}$ (red). We use the $315\, \mathrm{mK}$ (red) data to
extract the lever arm relating voltage along the red diagonal (see
Fig.~3(a)) to detuning $\epsilon$. At this temperature, electrons
are well thermalized to the refrigerator, and thus  $T_e \approx
T$. The width of the sensing transition  at this highest
temperature lets us extract the lever arm, which we then use to
estimate the electron temperature for the blue (green) data,
getting $T_e = 102(196)\, \mathrm{mK}$.

We next investigate the dependence of the sensing transition on
interdot tunneling in the regime of strong tunneling, $t \gtrsim
k_B T_e$. Figure~4 shows the left QPC sensing signal, again in
units of excess charge, along the detuning diagonal crossing a
different pair of triple points, at base temperature and for
various voltages on the coupling gate 11. For the weakest interdot
tunneling shown ($V_{11} = -1096$ mV), the transition was
thermally broadened, i.e., consistent with $t=0$ in Eqs.~(1) and
(2), and did not become narrower when $V_{11}$ was made more
negative. On the other hand, when $V_{11}$ was made less negative,
the transition widened as the tunneling between dots increased.
Taking $T_e = 102\, \mathrm{mK}$ for all data in Fig.~4 and calibrating
voltage along the detuning diagonal by setting $t=0$ for the
$V_{11} = -1096$ mV trace allows tunnel couplings $t$ to be
extracted from fits to our model of the other tunnel-broadened
traces. We find $t=10\, \mu\mathrm{eV}$ $(2.4\, \mathrm{GHz})$
(green trace), $t=16\, \mu\mathrm{eV}$ $(3.9\, \mathrm{GHz})$
(turquoise trace), and $t=22\, \mu\mathrm{eV}$ $(5.3\,
\mathrm{GHz})$ (purple trace). Again, fits to the two-level model
are quite good, as seen in Fig.~4.

Finally, we compare tunnel coupling values extracted from charge
sensing to values found using a transport-based method that takes
advantage of the $t$ dependence of the splitting of triple points
(honeycomb vertices) \cite{Ziegler00, vanderWiel03}. In the weak
tunneling regime, $t \ll \Delta$, the splitting of triple points
along the line separating isocharge regions $(M+1,N)$ and
$(M,N+1)$ has two components in the plane of gate voltages,
denoted here  $\delta V_{10}$ and $\delta V_{2}$. The lower and
upper triple points are found where the lowest energy $M+N+1$
state (the delocalized antisymmetric state) becomes degenerate
with the charge states $(M,N)$ and $(M+1,N+1)$, respectively.
Using the electrostatic model in Ref.~\cite{vanderWiel03}, we can
show that $\delta V_{10(2)}$ are related to various dot
capacitances and $t$ by
\begin{equation}
\delta V_{10(2)} =
\frac{|e|}{C_{g10(g2)}}\left(\frac{C_m}{C_o+C_m}+2t
\frac{C_o-C_m}{e^2}\right).
\end{equation}
Here, $C_{g10(g2)}$ is the capacitance from gate $10\, (2)$ to the
left (right) dot, $C_o$ is the self-capacitance of each dot, and
$C_{m}$ is the interdot mutual capacitance. All these capacitances
must be known to allow extraction of $t$ from $\delta V_{10(2)}$.
Gate capacitances $C_{g10(g2)}$ are estimated from honeycomb
periods along respective gate voltage axes, $\Delta V_{10(2)} \sim
|e|/C_{g10(g2)}\sim 6.8\, \mathrm{mV}$. Self-capacitances $C_o$
can be obtained from double dot transport measurements at finite
bias \cite{vanderWiel03}. However, lacking that data, we estimate
$C_o$ from single-dot measurements of Coulomb diamonds
\cite{LK97}. Mutual capacitance $C_m$ is extracted from the
dimensionless splitting $\delta V_{10(2)} /\Delta V_{10(2)} \sim
\frac{C_m}{C_o+C_m}\sim 0.2$, measured at the lowest tunnel
coupling setting.

Tunnel coupling values as a function of voltage on gate 11,
extracted both from charge sensing and triple-point separation,
are compared in the inset of Fig.~4. The two approaches are in
good agreement, with the charge-sensing approach giving
significantly smaller uncertainty for $t\gtrsim 0.5 k_B T_e$. The
two main sources of error in the sensing approach are uncertainty
in the fits (dominant at low $t$) and uncertainty in the lever arm
due to a conservative 10 percent uncertainty in $T_e$ at base.
Error bars in the transport method are set by the smearing and
deformation of triple points as a result of finite interdot
coupling and cotunneling. We note that besides being more
sensitive, the charge-sensing method for measuring $t$ works when
the double dot is fully decoupled from its leads. Like the
transport method, however, the sensing approach assumes $t\ll
\Delta$ (which may not be amply satisfied for the highest
values of $V_{11}$).

In conclusion, we have demonstrated differential charge sensing in
a double quantum dot using paired quantum point contact charge
sensors. States $(M+1,N)$ and $(M,N+1)$, with fixed total charge,
are readily resolved by the sensors, and serve as a two-level
system with a splitting of left/right states controlled by
gate-defined tunneling.  A model of local charge sensing of a
thermally occupied two-level system agrees well with the data.
Finally, the width of the $(M+1,N)\rightarrow(M,N+1)$ transition
measured with this sensing technique can be used to extract the
tunnel coupling with high accuracy in the range $0.5 k_B
T_e\lesssim t < \Delta$.

We thank M. Lukin, B. Halperin and W. van der Wiel for
discussions, and N. Craig for experimental assistance. We
acknowledge support by the ARO under DAAD55-98-1-0270 and DAAD19-02-1-0070, DARPA under the QuIST program, the NSF under DMR-0072777 and the Harvard NSEC,
Lucent Technologies (HJL), and the Hertz Foundation (LIC).

\end{document}